

\magnification=1200
\hsize = 5.5 true in
\hoffset = 0.5 true in
\baselineskip=20pt
\lineskiplimit=1pt
\lineskip=2pt plus .5pt
\parskip = 3pt
\font\tenrm=cmr10
\font\ninerm=cmr9

\def\today{\number\month /\number\day /\number\year}
\pageno=0
\headline={\rm \ifnum \pageno<1 LA-UR-94-2668 \hfill \today
		\else LA-UR-94-2668\hfil
``Water Ion Product K$_W$'' \hfil \today \fi}
\footline={\rm \ifnum \pageno<1 \hfil \else \hfil \folio \hfil \fi}

\topskip = 48pt
\parindent =  20pt
\vfil
\centerline{\bf Theoretical Calculation of the Water Ion}
\centerline{\bf Product K$_W$}

\bigskip

\centerline{Gregory J. Tawa and Lawrence R. Pratt}
\centerline{Theoretical Division }
\centerline{Los Alamos National Laboratory }
\centerline{Los Alamos, NM 87545}

\bigskip
\bigskip


\parindent =  20pt
\topskip = 10pt

\baselineskip=11pt
\ninerm

\beginsection{Abstract}

A dielectric solvation model is applied to the prediction of the
equilibrium ionization of liquid water over a wide range of density
and temperature with the objective of calibrating that model for the
study of ionization in water of organic acids, {\it e.g.\/}, proteins
and nucleic acids. The model includes an approximate description of
the polarizability of the dissociating water molecule.  The calculated
pK$_W$ are very sensitive to the value of the radii that parameterize
the model.  The radii required for the spherical molecular volumes of
the water molecule in order to fit the experimental ion product are
presented and discussed.  These radii are larger than those commonly
used.  They decrease with increasing density as would be guessed but
the rate of decrease is slight.  They increase with increasing
temperature, a variation opposite to what would be guessed if radii
were strictly viewed as a distance of closest approach.  The molecular
theoretical principles that might provide an explanation of the
thermodynamic state dependence of these radii are discussed.

\tenrm \baselineskip=20pt \bigskip \bigskip

\noindent {\bf 1. Introduction}

The chemical reaction H$_2$O = HO$^- + $ H$^+$ is of central
importance to a variety of practical problems including those
associated with chemical processes in aggressive aqueous environments
such as supercritical water [1, 2].  It has been considered the most
important chemical reaction in aqueous solution [3].

The equilibrium point of this process is often discussed on the basis
of the ion product K$_W$~=~[HO$^-$][H$^+$].  Measurements of K$_W$ for
water at densities below about twice the critical density are not
available though convincing model estimates of this property have been
given for the higher temperatures and pressures characteristic of
supercritical water processes [4, 5]. Moderately high temperatures are
required in order to examine K$_W$ over a continuous and extended
domain of thermodynamic states, including densities lower than those
liquid densities most commonly encountered.

This report gives theoretical calculations of K$_W$ over a wide range
of thermodynamic states.  The calculations here employ a dielectric
solvation model similar to that used by Pitzer and collaborators [4,
5].  As an `equation of state' model the present results are not
expected to improve upon that prior work.  But the present
contribution is helpful for several reasons that focus on the basis of
the theory.  The model used here is simple and the connection with
molecular theory is simple too [6-10]. Thus the results here will
facilitate direct molecular-scale verification of a number of
intermediate quantities.  This topic is of recent theoretical interest
[11-14]; we expect comparable molecular-scale theory and results to be
available in the near future.

Since the model we exploit is common in studies of the charge states
of protein and nucleic acid molecules in aqueous solutions, the
molecular scale verifications are expected to improve our
understanding of those biophysical problems too [15-23].

A concern with dielectric solvation models is the assignment of cavity
radii or volumes.  Cavity volumes should depend on the thermodynamic
state [8, 10, 24, 25]. This is important to the extent that
derivatives of the solvation free energy with respect to thermodynamic
state, {\it i.e.\/} temperature, pressure, and composition, are
required. The molecular theoretical principles upon which those radii
might be determined have been identified [7, 8, 10]. However, this
issue can also be studied empirically by analyzing the solvation free
energy computed on the basis of the dielectric solvation model over a
sufficiently wide range of thermodynamic conditions.  The data on the
chemical equilibrium that is studied in this report are more extensive
than for most reactions in aqueous solutions.  The water
self-ionization reaction is thus suitable for the principal motivation
of this work: the delineation of the region where the model works
satisfactorily and a calibration of the radii required to describe
ionization of organic acids.

The next section gives technical details of the theoretical issues and
methods.  We then present the results and a discussion.

\bigskip \noindent{\bf 2. Methods}

The dielectric solvation model used here is formulated from the
following physical ingredients. Attention is focused on a solute of
interest and a solute volume is defined.  For liquid water under the
most common conditions it is known that the van der Waals volume of
the molecule is a reasonable initial choice for the molecular volume.
Partial charges describing the solute electric charge distribution are
positioned within this volume.  The solvent is idealized as a
continuous dielectric material with the measured dielectric constant
$\varepsilon$.  The solvent is considered to be excluded from the
solute volume and that molecular volume is assigned an internal
dielectric constant $\varepsilon_m$; when $\varepsilon_m > 1$ the
model thereby includes an approximate representation of the molecular
polarizability [26].  In all the calculations below we have used
$\varepsilon_m = 2.42$ which corresponds approximately to the known
polarizability of the water molecule (1.44 \AA$^3$) when a spherical
molecular volume of radius 1.65 \AA \ is assumed.  Note, however, that
when adjustments of the radii are considered below the value for
$\varepsilon_m$ is not adjusted.

The equation to be solved for the model is $$\nabla \bullet
\varepsilon ({\bf r})\nabla \Phi ({\bf r})=-\;4\pi \rho _f({\bf r}) ,
\eqno{(1)}$$ where $\rho_f ({\bf r})$ is the density of electric
charge of the solute mole\-cule, the function $\varepsilon({\bf r})$
gives the local value of the dielectric constant, and the solution
$\Phi({\bf r})$ is the electric potential.  To solve this equation, we
first cast it as an integral equation, {\it e.g.\/}, $$\Phi ({\bf
r})=\Phi ^{(0)}({\bf r}) / \varepsilon_m +\int\limits_V {G^{(0)}({\bf
r},{\bf r}')\left( {{{\nabla '\varepsilon ({\bf r}')} \over {4\pi
\varepsilon ({\bf r}')}}} \right)\bullet \nabla '\Phi ({\bf
r}')d^3r'}.  \eqno{(2)} $$ Here $G^{(0)}({\bf r},{\bf r}')$ is the
Green function for the Poisson equation with $\varepsilon ({\bf r})=1$
and $\Phi ^{(0)}({\bf r})$ is the electrostatic potential for that
case.  This equation is correct both for a localized distribution
$\rho_f ({\bf r})$ and zero boundary data on a surface everywhere
distant and for periodic boundary conditions on a cell of volume $V$.
$G^{(0)}({\bf r},{\bf r}')$ is different in those two cases as is
$\Phi ^{(0)}({\bf r})$.  For the unbounded cases treated here
$G^{(0)}({\bf r}, {\bf r}') = 1/\vert{\bf r} - {\bf r}' \vert$.  This
equation is not the only such form that can be solved but it is the
basis for the most popular boundary element approaches for numerical
solution of this physical model [27-33].  Because of this we defer a
more detail discussion of integral formulations of Eq.~1.

The integrand of Eq.~2 is concentrated on the interface between the
solute volume and the solvent.  The interpretation of the integral of
Eq.~2 is that it superposes contributions to the electrostatic
potential due to charge induced on the surface.  We can then use
boundary element ideas to solve it.  We have used sampling methods,
here principally those based upon quasi-random number series [34-36],
to evaluate the surface integral rather than more specialized methods.
Advantages of such approaches are that they are simple, yet permit
systematic studies of numerical convergence and exploitation of
systematic coarse-graining.  The geometries considered here are
particularly simple, so these applications are not a stringent test of
numerical methods.  But an elegant statement of the advantages of
boundary element approaches in the this context can be found in a
paper of Yoon and Lenhoff [30].

In this application we solved Eq.~2 for the molecule in the liquid
with the experimental dielectric constant and in isolation with the
external dielectric constant assigned a value of $\varepsilon = 1$.
The solutions for these cases are denoted by $\Phi_l({\bf r})$ and
$\Phi_v({\bf r})$, respectively.  Remember that the `internal'
dielectric constant parameter is fixed at $\varepsilon_m=2.42$
throughout.  Then we construct $$\Delta \mu =\left( {{1 \over 2}}
\right)\int\limits_V {\rho _f({\bf r})\left( {\Phi_l ({\bf r})-\Phi
_v({\bf r})} \right)d^3r}, \eqno{(3)} $$ the standard chemical
potential difference used below.  This difference involves only the
indirect contributions of Eq.~2, {\it i.e.\/} an infinite self-energy
is eliminated by this subtraction.  Here $\rho_f ({\bf r})$ is a sum
of partial charges. Therefore the integral in Eq.~3 is a sum over
those partial charges.

With applications of such an {\it ad hoc\/} model, some arbitrariness
is practically unavoidable.  For these calculations the most important
features of the method that are arbitrary are two: (i) the chemical
equilibrium and chemical species that are treated; and (ii) the
intramolecular volumes assigned.

The chemical reaction that we study in order to calculate K$_W$ is $$2
\ H_2O = H_3O^+ + HO^-.  \eqno{(4)}$$ This arbitrary choice is made on
the basis of the physical guess that observation of a `free' proton
would be less likely than the case where an excess proton is
associated with an oxygen atom in roughly the same circumstances that
an H-O atom pair are configured in the water molecule.  Atoms more
distant from an `excess' proton than the nearest oxygen should be of
secondary importance.  We then consider the equilibrium ratio $$K =
{{\left[ {H_3O^+} \right]\left[ {HO^-} \right]} \over {\left[ {H_2O}
\right]^2}}. \eqno{(5)}$$ We express this quantity as $$K = K^{(0)}(T)
\times \exp \left\{ - \left( \Delta\mu_{HO^-} +\Delta\mu_{H_3O^+} -
2\Delta\mu_{H_2O} \right)/RT \right\},\eqno{(6)}$$ where $K^{(0)}(T)$
is the equilibrium constant in the absense of intermolecular
interactions, {\it i.e.\/} for the ideal gas [11, 12]. Accounting for
the mass balance of the Eq.~4 we then find $$K_W \equiv [H_3O^+][HO^-]
= \rho^2 K \left[ 1-2 K^{1/2} \over 1-4 K \right]^2, \eqno{(7)}$$
where $\rho$ is the formal density of the water.

The ideal gas contribution $K^{(0)}(T)$ can be obtained from standard
formulae and information on the molecular structures of the species
considered.  The information used here is collected in Table I
[37-39].

What remains is the evaluation of the excess chemical potentials
appearing in Eq.~6 for the equilibrium ratio K.  This brings us to the
second arbitrary feature of our calculation, the assignment of
molecular volumes.  Here we assumed that each molecular volume is a
sphere centered on the oxygen atom and that the radii of the spheres
for all species were equal.  This constraint is not necessary.
However, some fitting to experimental results will be required and
limiting the number of fitting parameters seems prudent.  It might be
hoped that the fitted radii would validly describe ionization of
organic acids too; if the radii of protonated and deprotonated oxygen
atoms were not transferrable to different but similar chemical
environments with acceptable accuracy much of the usefulness of the
model would be lost.  In addition, the molecular underpinnings of the
dielectric solvation model have been clarified over recent years; for
solute species with spherical short-ranged (non-electrostatic)
interactions with the solvent, we can hope that sufficient molecular
information will become available to make applications such as this
more geniunely predictive [7, 8, 10].

We note the precedent for cations to be assigned slightly larger radii
than the isoelectronic anions [40, 41]; several explanations of this
have been offered [8, 40, 41]. Here we assume that such complications
are sufficiently slight that we can ignore them.

Finally we must specify the charge distribution $\rho_f({\bf r})$ of
the chemical species treated and the dielectric constant.  For the
former quantity we used atom centered partial charges following the
method of Breneman and Wiberg [42]. The dielectric constant utlized
was that given by the empirical equation of Quist and Marshall [43].

\bigskip \noindent{\bf 3. Results and Discussion}

Our results for K$_W$ are shown in Fig.~1.  The densities treated
range over a factor of three including the triple point density.  The
lowest density shown is approximately 0.4 g/cm$^3$.  Temperatures
encompass all of the traditional liquid range from the triple
temperature to above the critical temperature.  Fig.~2 shows the
empirical radii that resulted from fitting the model result to
experimental values given by Marshall and Franck [3].  The magnitudes
of these radii are reasonable but they are significantly larger than
would be most commonly guessed [41]. For the upper part of this
density range, the radii are decreasing functions of density at
constant temperature.  But the variations of the radii with density at
fixed temperatures are slight; the decreases in the radii are 3-7
times smaller than the rule $\delta R / R \approx - \delta \rho /3
\rho$ based upon the Onsager estimate of the radius, $4 \pi R^3 \rho
/3 = 1$ [24, 44].  The variation observed for R with temperature at
fixed density is not accounted for by the Onsager radius.  Moreover, a
natural guess for the variation of R with temperature at constant
density is not so clear; if R were identified as a distance of closest
approach of the solvent molecules to the solute then R would be
expected to decrease with increasing temperature at fixed density.

The sensititivity to the parameter R is an important issue for
applications of the model.  In the present case, pK$_W$ can change by
1 unit if the value of R is changed by 1\% from the value that fits
the experiment.  A change of 2\% in R from the optimum value can lead
to a change of more than 3 units in pK$_W$.  Computational searches
for a set of three radii $(R_{H_2O}, R_{H_3O^+}, R_{HO^-})$ that could
satisfactorily fit the whole behavior in Fig.~1 without thermodynamic
state dependence were unsuccessful.  For example consider the isotherm
at $T=0$ C of Fig.~1 for which the total change in pK$_W$ is about
1.7.  The set of three radii that we found to give the best fit to
those data still lead to errors of +0.57 and -0.55 at the beginning
and the end, respectively, of the isotherm shown.  The set of three
radii that produced the best fit to the data over the whole of Fig.~1
lead to typical errors in pK$_W$ greater than 4 and those radii were
not chemically reasonable.

How the model corresponds most naturally to molecular theory is known
[7, 8, 10] and this correspondence provides an avenue for theoretical
determination of R.  The corresponding molecular theory is the
second-order cumulant approximation $$ \Delta \mu \approx \Delta \mu
_0 + \left< \sum_j \varphi({\bf j}) \right> _0 -{\beta \over 2} \left<
{\left( \sum_j \varphi ({\bf j}) - \left< \sum_k \varphi ({\bf k})
\right> _0 \right)}^2 \right> _0 .  \eqno{(8)}$$ Here $ \varphi({\bf
j}) $ is the electrostatic interaction potential energy coupling the
solute to solvent molecule {\bf j}.  The brackets $\left< \cdots
\right>_0$ indicate the thermal average in the absence of those
electrostatic couplings and $\Delta \mu_0$ is the excess chemical
potential of the solute at infinite dilution again in the absence of
electrostatic interactions.  When the solute is an infinitely dilute
second component this molecular approximation is perturbation theory
through second order in the electrostatic interactions.  When the
molecule is not literally an infinitely dilute second component the
`infinite dilution' restriction means that one molecule is
distinguished for the purposes of calculation.  This is still a
natural theory but the medium now contains a non-zero concentration of
molecules mechanically identical to the `solute.' The medium
properties non-perturbatively reflect that fact.  From the perspective
of this molecular theory, the dielectric solvation model application
neglects the zeroth and first order terms, makes an estimate of the
second-order term, and neglects all succeeding contributions.

Careful calculation on a molecular basis of the terms in Eq.~8 would
be a helpful next step for clarification of these theoretical models.
Jayaram, {\it et al.\/} [45] and Rick and Berne [46] have tested
dielectric solvation models by computer experiment.  Their results
provide, in principle, the information requested here but they were
not analyzed from the point of view of the present goals for this
system.  Those results together suggest that hydrogen-bonding
interactions are border-line cases for dielectric solvation models.
Although the present application and its motivation in ionization of
organic acids is of wide interest, it is likely a severe test for
these models.  Thus the empirical radius R determined here probably
subsumes other errors that would have been encountered had the
molecular theory been implemented.

\noindent {\bf Acknowledgements}

We are grateful to Drs. J. Blair, S.-H.  Chou, P.  Leung, D. Misemer,
J. Stevens, and K. Zaklika of 3M Corporation for helpful discussions
on topics of reaction chemistry in solution.  LRP thanks Gerhard
Hummer and Angel E. Garcia for helpful discussions and acknowledges
partial support for this work by the US-DOE under LANL Laboratory
Directed Research and Development funds.  Andrew Pohorille made useful
comments on an earlier version of this report.  This work was also
supported in part Tank Waste Remediation System (TWRS) Technology
Application program, under the sponsorship of the U. S.  Department of
Energy EM-36, Hanford Program Office, and the Air Force Civil
Engineering Support Agency.

\vfil \eject \noindent {\bf References}

\item{1.} {\it Chemical and Engineering News\/}, December 23, 1991,
pp.  26-39: ``Supercritical water.  A medium for chemistry.''

\item{2.} Marshall, W. L.; Franck, E. U. {\it J. Phys. Chem.  Ref.
Data\/} {\bf 1981}, 10, 295.

\item{3.} Stillinger, F. H. {\it Theoretical Chemistry, Advances and
Perspectives\/}; edit\-ed by H. Eyring and D. Henderson; Academic: NY,
1978, vol. 3, pp. 177-234.

\item{4.} Pitzer, K. S. {\it J. Phys. Chem.\/} {\bf 1982}, 86, 4704.

\item{5.} Tanger, J. C. IV; Pitzer, K. S. {\it AIChE Journal\/} {\bf
1989}, 35, 1631.

\item{6.} Hirata, F.; Redfern, P.; Levy, R. M. {\it Int. J. Quant.
Chem.\/} {\bf 1988}, 15, 179.

\item{7.} Levy, R. M.; Belhadj, M.; Kitchen, D. B. {\it J.  Chem.
Phys.\/} {\bf 1991}, 95, 3627.

\item{8.} Pratt, L. R.; Hummer, G.; Garcia, A. E. {\it Biophys.
Chem.\/} {\bf 1994}, 51, 147.

\item{9.} Figueirido, F.; Del Buono, G. S.; Levy, R. M. {\it
Biophys. Chem.\/} {\bf 1994}, 51, 235.

\item{10.} Tawa, G. J.; Pratt, L. R. {\it Structure and reactivity in
aqueous solution. Characterization of chemical and biological
systems\/}; edited by C. J.  Cramer and D. G. Truhlar; American
Chemical Society, Washington DC; 1994, vol. 586, pp. 60-70.

\item{11.} Guissani, Y.; Guillot, B.; Bratos, S. {\it J. Chem.
Phys.\/} {\bf 1988}, 88, 5850.

\item{12.} Nyberg, A.; Haymet, A. D. J. {\it Structure and reactivity
in aqueous solution. Characterization of chemical and biological
systems\/}; edited by C. J.  Cramer and D. G. Truhlar; American
Chemical Society, Washington DC; 1994, vol. 586, pp. 111-119.

\item{13.} Ando, K.; Hynes, J. T. {\it Structure and reactivity in
aqueous solution. Characterization of chemical and biological
systems\/}; edited by C. J.  Cramer and D. G. Truhlar; American
Chemical Society, Washington DC; 1994, vol. 586, pp. 143-153.

\item{14.} Jorgensen, W. L.; Briggs, J. M. {\it J. Am. Chem. Soc.\/}
{\bf 1989}, 111, 4190.

\item{15.} Bashford, D.; Karplus, M. {\it Biochem.\/} {\bf 1990}, 29,
10219.

\item{16.} Bashford, D.; Karplus, M. {\it J. Phys. Chem.\/} {\bf
1991}, 95, 9556.

\item{17.} Merz, K. B., Jr. {\it J. Am. Chem. Soc.\/} {\bf 1991},
113, 3572.

\item{18.} Honig, B.; Sharp, K.; Yang, A.-S. {\it J. Phys. Chem.\/}
{\bf 1993}, 97, 1101.

\item{19.} Yang, A.-S.; Honig, B. {\it J. Mol. Biol.\/} {\bf 1993},
231, 459.

\item{20.} Yang, A.-S.; Gunner, M. R.; Sampogna, R.; Sharp, K.; and
Honig, B. {\it Proteins\/} {\bf 1993}, 15, 252.

\item{21.} Oberoi, H.; Allewell, N. M. {\it Biophys. J.\/} {\bf
1993}, 65, 48.

\item{22.} Young, P.; Green, D. V. S.; Hillier, I. H.; Burton, N. A.
{\it Molec. Phys.\/} {\bf 1993}, 80, 503.

\item{23.} Del Buono, G. S.; Figueirido, F.; Levy, R. M. {\it
Proteins\/} {\bf 1994}, 20, 85.

\item{24.} B\"ottcher, C. J. F. {\it Theory of Electric
Polarization\/}; Elsevier: NY, 1973, 2nd edition, volume 1, chapter 4.

\item{25.} Roux, B.; Yu, H.-A.; Karplus, M. {\it J. Phys. Chem.\/}
{\bf 1990}, 94, 4683.

\item{26.} Sharp, K.; Jean-Charles, A.; Honig, B. {\it J. Phys.
Chem.\/} {\bf 1992}, 96, 3822.

\item{27.} Pascual-Ahuir, J. L.; Silla, E.; Tomasi, J.; Bonaccorsi, R.
{\it J. Comp. Chem.\/} {\bf 1987}, 8, 778.

\item{28.} Zauhar, R. J.; Morgan, R. S. {\it J. Mol. Biol.\/} {\bf
1985}, 186, 815; {\it J. Comp.  Chem.\/} {\bf 1988}, 9, 171.

\item{29.} Rashin, A. A.; Namboodiri, K. {\it J. Phys. Chem.\/} {\bf
1987}, 91, 6003.

\item{30.} Yoon, B. J.; Lenhoff, A. M. {\it J. Comp. Chem.\/} {\bf
1990}, 11, 1080; {\it J. Phys. Chem.\/} {\bf 1992}, 96, 3130.

\item{31.} Juffer, A. H.; Botta, E. F. F.; van Keulen, A. M.; van der
Ploeg, A.; Berendsen, H. J. C. {\it J. Comp. Phys.\/} {\bf 1991}, 97,
144.

\item{32.} Wang, B.; Ford, G. P. {\it J. Chem. Phys.\/} {\bf 1992},
97, 4162.

\item{33.} Zhou, H.-X. {\it Biophys. J.\/} {\bf 1993}, 65, 955.

\item{34.} Hammersley, J. M.; Handscomb, D. C. {\it Monte Carlo
Methods\/}; Chapman and Hall: London, 1964, pp. 31-36.

\item{35.} Press, W. H.; Teukolsky, S. A.; Vetterling, W. T.;
Flannery, B. P. {\it Numerical Recipes, the Art of Scientific
Computing\/}; Cambridge University Press: NY, 1992, 2nd edition, \S
7.7.

\item{36.} Niederreiter, H. {\it Random Number Generation and
Quasi-Monte Carlo Methods\/}; SIAM: Philadelphia, 1992.

\item{37.} McQuarrie, D. A. {\it Statistical Mechanics\/}; Harper and
Row: New York, 1976, Chapter 8.

\item{38.} Kern, C. W.; Karplus, M. in {\it Water: A Comprehensive
Treatise\/}; edited by F.Franks; Plenum Press: New York, 1972, Chapter
2.

\item{39.} Stillinger, F.H.;  David, C. W. {\it J. Chem. Phys.\/}
{\bf 1978}, 69,  1473.

\item{40.} Latimer, W. M.; Pitzer, K. S.; Slansky, C. M. {\it J.
Chem. Phys.\/} {\bf 1939}, 7, 108.

\item{41.} Rashin, A. A.; Honig, B. {\it J. Phys. Chem.\/} {\bf
1985}, 89, 5588.

\item{42.} Breneman, C. M.; Wiberg, K. B. {\it J. Comp. Chem.\/} {\bf
1990}, 11, 361.

\item{43.} Quist, A. S.; Marshall, W. L. {\it J. Phys. Chem.\/} {\bf
1965}, 69, 3165.

\item{44.} Onsager, L. {\it J. Am. Chem. Soc.\/} {\bf 1936}, 58,
1486.

\item{45.} Jayaram, B.; Fine, R.; Sharp, K.; Honig, B. {\it J. Phys.
Chem.\/} {\bf 1989}, 93, 4320.

\item{46.} Rick, S. W.; Berne, B. J. {\it J. Am. Chem. Soc.\/} {\bf
1994}, 116, 3949.

\vfill \eject

\noindent {\bf Figure Captions}

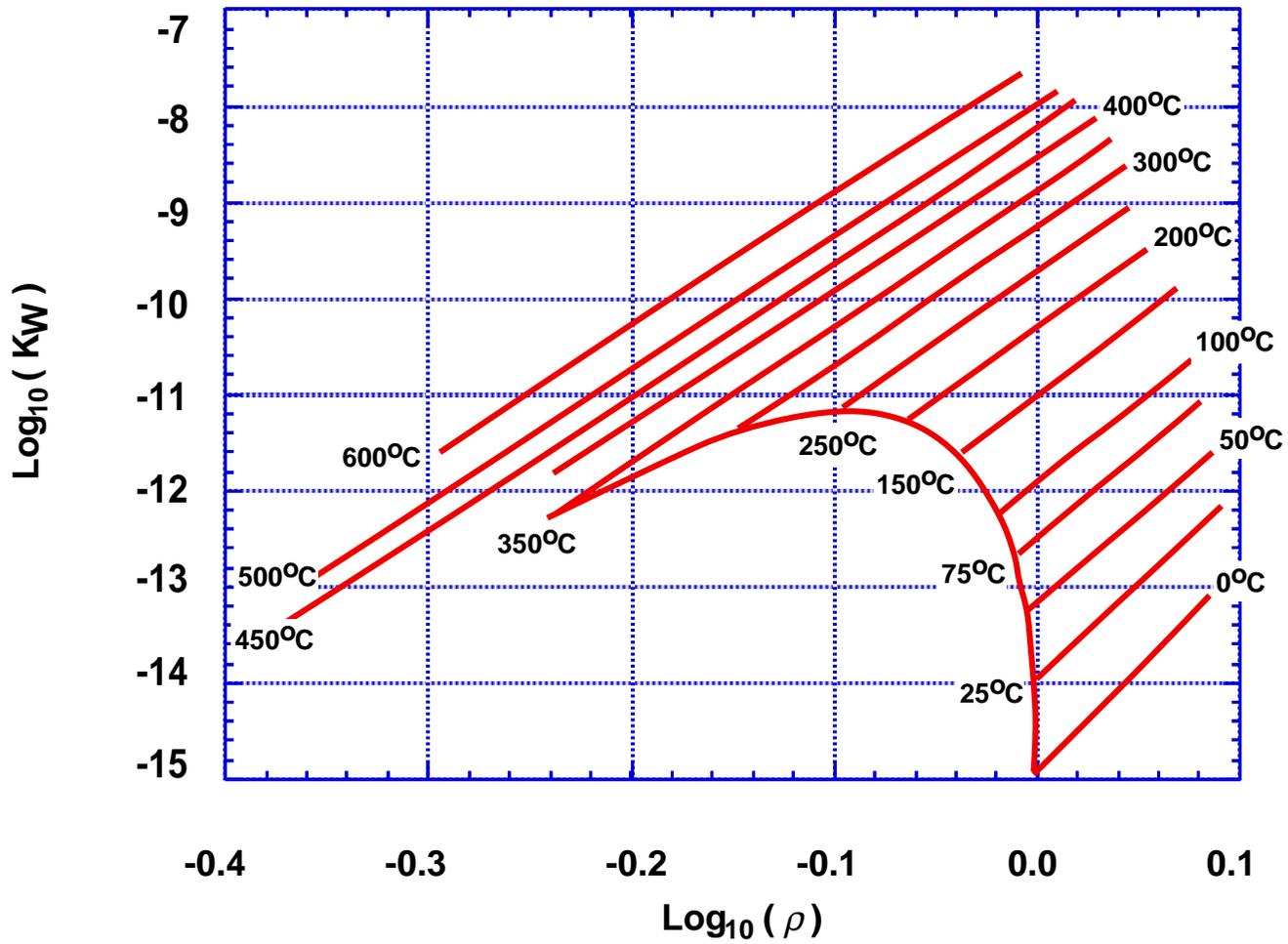
\item{Figure 1:} The water ion product K$_W$ ([mol/l]$^2$) as a
function of density $\rho$ (g/cm$^3$) and temperature over a range of
liquid thermodynamic states.  The region underneath the domed-shaped
curve is the liquid-vapor coexistence region.

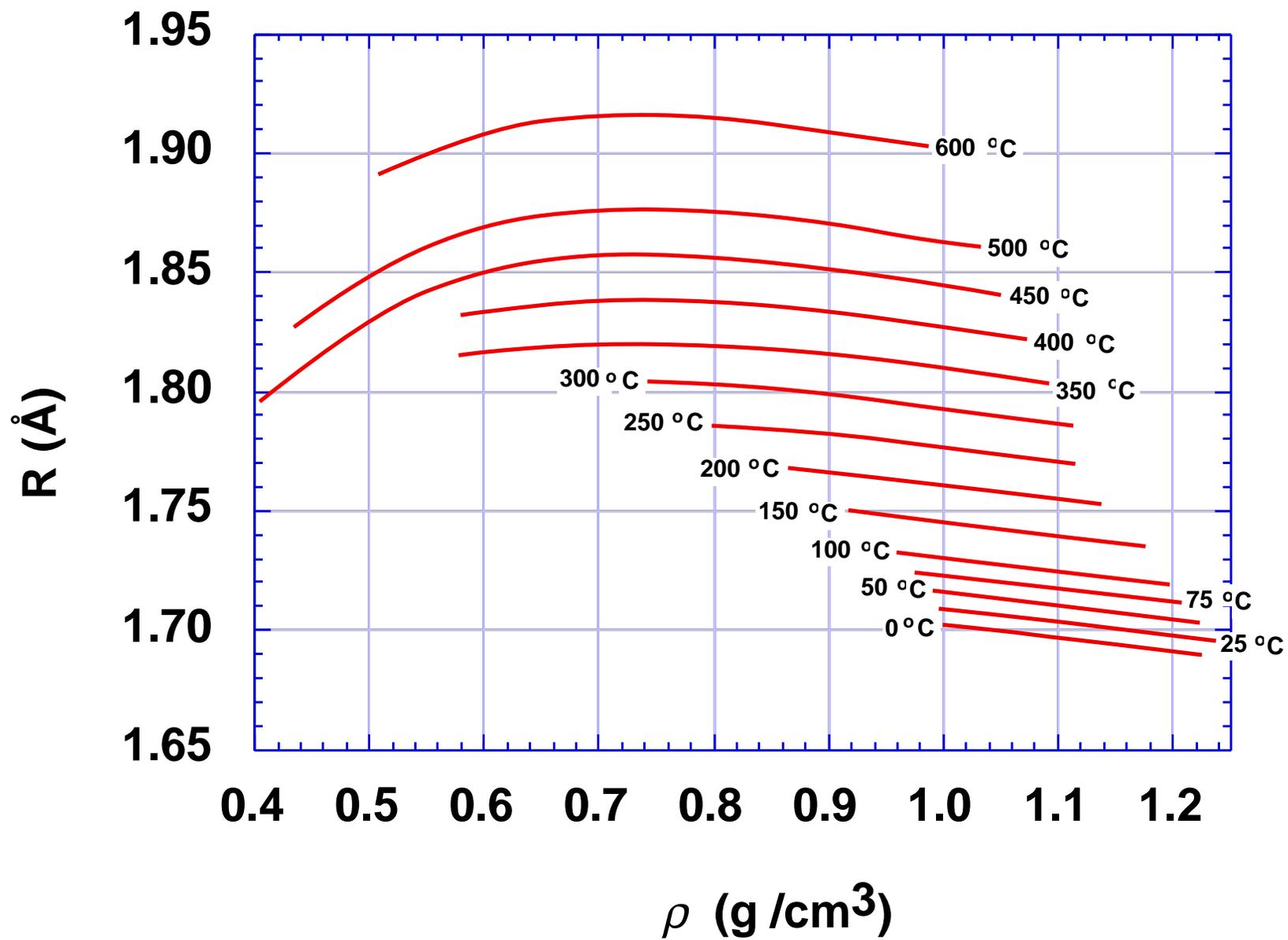
\item{Figure 2:} Empirical radii R, as a function of density and
temperature, required for the dielectric solvation  model to fit the
experimental results.

\vfill \eject

\noindent {\bf Table I.  Properties used in the calculation of
molecular partition functions} \bigskip

\vrule{\offinterlineskip
\halign{&\vrule#&\strut\quad\hfil\strut\quad\hfil
	\strut\quad\hfil#\quad\cr
\noalign{\hrule}
height2pt&\omit&&\omit&&\omit&&\omit&\cr
&property\hfill&&H$_2$O\hfil$^a$&&H$_3$O$^+$\hfil$^b$&&HO$^-$\hfil$^b$&\cr
height2pt&\omit&&\omit&&\omit&&\omit&\cr
\noalign{\hrule}
height2pt&\omit&&\omit&&\omit&&\omit&\cr
&$\sigma$\hfill$^c$&&2&&1&&1&\cr
&$\Theta_{v1}$\hfill$^d$&&2679.48&&2067.21&&7603.99&\cr
&$\Theta_{v2}$\hfill&&5641.60&&2694.40&&---\hfil&\cr
&$\Theta_{v3}$\hfill&&5796.13&&2736.17&&---\hfil&\cr
&$\Theta_{v4}$\hfill&&---\hfil&&4267.23&&---\hfil&\cr
&$\Theta_{v5}$\hfill&&---\hfil&&4304.29&&---\hfil&\cr
&$\Theta_{v6}$\hfill&&---\hfil&&4315.57&&---\hfil&\cr
&$\Theta_{A}$\hfill$^e$&&13.6745&&8.3841&&33.9547&\cr
&$\Theta_{B}$\hfill&&20.9686&&13.4501&&---\hfil&\cr
&$\Theta_{C}$\hfill&&39.1187&&13.9630&&---\hfil&\cr
&E$_g$\hfill$^f$&&$-$1032.98&&$-$1203.55&&$-$643.127&\cr
height2pt&\omit&&\omit&&\omit&&\omit&\cr
\noalign{\hrule}
}
}

\hfill
\bigskip \item{ $^a$} Geometry from Ref. 38.

\item{ $^b$} Geometry from Ref. 39.

\item{$^c$} Rotational symmetry number.

\item{ $^d$} Vibrational temperatures as defined in Ref. 37, in
kelvin.  These properties were calculated using SCF approximation, a
6-31g* basis set, and the geometries of Refs. 38 and 39.

\item{ $^e$} Rotational temperatures as defined in Ref. 37, in kelvin.
These properties were calculated in the same way as the vibrational
temperatures were.

\item{ $^f$} Electronic ground state energies from Ref. 39 relative
to the energies of the separated isolate atoms, in
kcal/mol.

\end